\documentstyle[12pt, epsfig]{article}  
\begin{document} 
\baselineskip=18pt

\def\la{\mathrel{\mathpalette\fun <}}
\def\ga{\mathrel{\mathpalette\fun >}}
\def\fun#1#2{\lower3.6pt\vbox{\baselineskip0pt\lineskip.9pt
\ialign{$\mathsurround=0pt#1\hfil##\hfil$\crcr#2\crcr\sim\crcr}}}  
\def\lrang#1{\left\langle#1\right\rangle}

\begin{titlepage}   
\title{NUCLEAR GEOMETRY OF JET QUENCHING} 
\vskip 1cm 
\author{I.P.Lokhtin, A.M.Snigirev \\ ~ \\ 
\it M.V.Lomonosov Moscow State University, \\  
\it D.V.Skobeltsyn Institute of Nuclear Physics,\\  \it Moscow, Russia } 
\date{}
\maketitle 
\vskip 1 cm   
\begin{abstract} 
The most suitable way to study the jet quenching as a function of distance traversed
is varying the impact parameter $b$ of ultrarelativistic nucleus-nucleus collision 
(initial energy density in nuclear overlapping zone is almost independent of $b$ up to 
$b \sim R_A$). It is shown that $b$-dependences of medium-induced radiative and 
collisional energy losses of a hard parton jet propagating through dense QCD-matter are 
very different. The experimental verification of this phenomenon could be performed for 
a jet with non-zero cone size basing on essential difference between angular 
distributions of collisional and radiative energy losses.   
\end{abstract}
\end{titlepage}    
\newpage 

\section {Introduction}   

The experimental investigation of ultrarelativistic nuclear collisions offers a unique 
possibility of studying the properties of strongly interacting matter at the high 
energy density when the hadronic matter is expected to become deconfined and a gas of 
asymptotically free quarks and gluons is formed. This is called quark-gluon plasma 
(QGP), in which the colour interactions between partons are screened owing to 
collective effects (see, for example, 
reviews~\cite{satz182,muller96,lok_rev,bass_rev}).  

In recent years, a great deal of attention has been paid to the study of "hard"
probes of QGP -- heavy quarkonia and hard partonic jets, which do not appear as
constituents of the thermalized system, but can carry information about the
earliest stages of its evolution. In particular, the strong suppression of yield of 
heavy quark vector mesons as $J/\Psi$, $\Psi '$ ($c \bar{c}$ states) and $\Upsilon$, 
$\Upsilon '$, $\Upsilon ''$ ($b \bar{b}$ states) is one of the promising signatures of 
the quark-gluon plasma formation in heavy ion collisions~\cite{satz86}. An intriguing 
phenomenon is the "anomalously" small yield of $\Psi$-resonances, observed in Pb-Pb 
collisions in the NA50 experiment (CERN-SPS)~\cite{na50} and inconsistent with the 
conventional model of pre-resonance absorption in cold nuclear matter. Although the 
interpretation of this phenomenon as a result of the formation of a QGP is quite 
plausible~\cite{psi-qgp}, alternative explanations have also been put forward, such, 
for example, as $\Psi - h$ rescattering on comoving hadrons~\cite{psi-had}. Thus the 
nature of this "anomalous" suppression of $\Psi$-resonance production is not yet fully 
understood, and it should be completely explained in future~\cite{vogt-psi}. For 
heavier ($b \bar{b}$) systems, a similar suppression effect in super-dense strongly 
interacting matter is expected at higher temperatures than for $c \bar{c}$, which are 
expected to be reached in central collisions of heavy ions at the RHIC at BNL and LHC 
at CERN colliders. 

Along with the suppression of heavy quarkonia, one of the processes which may give
information about the earliest stages of evolution of the dense matter formed in
ultrarelativistic nuclear collisions is the passage through the matter of hard jets of 
colour-charged partons, pairs of which are created at the very beginning of the 
collision process (typically, at $\la 0.01$ fm/c) as a result of individual initial 
hard nucleon-nucleon (parton-parton) scatterings. Such jets pass through the dense 
parton matter formed due to mini-jet production at larger time scales 
($\sim 0.1$ fm/c), and interact strongly with the comoving constituents in the medium, 
changing its original properties as a result of additional rescatterings. The inclusive 
cross section for hard jet production processes is still very small for performing a 
systematic analysis at the SPS energies ($\sqrt{s} \simeq 20$ GeV per 
nucleon pair), but it increases fast with the energy of collided nuclei. Thus these 
will play important role in the formation of the initial state at the energies of  
RHIC ($\sqrt{s} = 200$ GeV per nucleon pair) and LHC ($\sqrt{s} = 5.5$ TeV per nucleon
pair) colliders. 
 
The actual problem is to study the energy losses of a hard jet evolving through the 
dense matter. We know two possible mechanisms of energy losses: $(1)$ radiative losses 
due to gluon "bremsstrahlung" induced by multiple 
scattering~\cite{ryskin,gyul94,baier,baier2,zakharov,urs99} and $(2)$ collisional 
losses due to the final state interactions (elastic rescatterings) of high $p_T$ 
partons off the medium constituents~\cite{bjork82,mrow91,lokhtin1}. Since the jet 
rescattering intensity strongly increases with temperature, formation of a super-dense 
and hot partonic matter in heavy ion collisions (with initial temperature up to $T_0 
\sim 1$ GeV at LHC~\cite{eskola94}) should result in significantly larger jet 
energy losses as compared with the case of "cold" nuclear matter or hadronic gas at 
$T \la 0.2$ GeV. 

Although the radiative energy losses of a high energy parton have been shown to 
dominate over the collisional losses by up to an order of magnitude~\cite{gyul94}, a 
direct experimental verification of this phenomenon remains an open problem. Indeed, 
with increasing of hard parton energy the maximum of the angular distribution of 
bremsstrahlung gluons has shift towards the parent parton direction. This means that 
measuring the jet energy as a sum of the energies of final hadrons moving inside an 
angular cone with a given finite size $\theta_0$ will allow the bulk of the gluon 
radiation to belong to the jet and thus the major fraction of the initial parton energy 
to be reconstructed. Therefore, the medium-induced radiation will, in the first place, 
soften particle energy distributions inside the jet, increase the multiplicity of 
secondary particles, but will not affect the total jet energy. It was recently 
shown~\cite{baier,baier2} that the radiation of energetic gluons in a QCD medium is 
essentially different from the Bethe-Heitler independent radiation pattern. Such gluons 
have formation times exceeding the mean free path for QCD parton scattering in the 
medium. In these circumstances the coherent effects play a crucial role leading to a 
strong suppression of the medium-induced gluon radiation. This coherent suppression is 
a QCD analogue of the Landau-Pomeranchuk-Migdal effect in QED. It is important to 
notice that the coherent LPM radiation induces a strong dependence of the jet energy on 
the jet cone size $\theta_0$~\cite{lokhtin2,baier3}.  

On the other hand, the collisional energy losses represent an incoherent sum over 
all rescatterings. It is almost independent of the initial parton energy. Meanwhile,  
the angular distribution of the collisional energy loss is essentially different 
from that of the radiative one. The bulk of "thermal" particles knocked out of the 
dense matter by elastic scatterings fly away in almost transverse direction relative to 
the hard jet axis. As a result, the collisional energy loss turns out to be practically 
independent on $\theta_0$ and emerges outside the narrow jet cone. Thus the relative 
contribution of collisional losses would likely become significant for jets with finite 
cone size propagating through the QGP~\cite{lokhtin2}. 

In a search for experimental evidences in favour of the medium-induced energy losses a 
significant dijet quenching (a suppression of high-$p_T$ jet pair yield)~\cite{gyul90} 
and a monojet-to-dijet ratio enhancement~\cite{gyulqm95} were proposed as possible 
signals of dense matter formation in ultrarelativistic collisions of nuclei. Other 
possible signatures that could directly measure the energy losses involve tagging the 
hard jet opposite a particle that does not interact strongly as a 
$Z$-boson~\cite{kvat95} (mostly $q + g \rightarrow q + Z (\rightarrow \mu^+ \mu^-)$, 
but also  $q + \bar{q} \rightarrow q + Z$ ) or a photon~\cite{wang96} (mostly 
$q + g \rightarrow q + \gamma$, also  $q + \bar{q} \rightarrow q + \gamma$). The jet 
energy losses in dense matter should result in the non-symmetric shape of the 
distribution of differences in $P_T$ between the Z-boson ($\gamma$) and jet. The above 
phenomena can be studied in heavy ion collisions~\cite{note99-016} with Compact Muon 
Solenoid (CMS), which is the general purpose detector designed to run at the 
LHC~\cite{cms94}. Note, that using $\gamma + jet$ channel in this case is complicated 
due to large background from $jet+jet$ production when one of the jet in an event is 
misidentified as a photon (the leading $\pi ^{0}$). However the shape of the 
distribution of differences in $E_T$ between the $\gamma$ and jet is very different for 
signal and background, and still sensitive to the jet energy losses~\cite{note99-016}. 

The advantage of $\gamma + jet$ and $Z(\rightarrow \mu^+\mu^-) + jet$ channels is that 
one can determine the average initial transverse momentum of the hard jet, 
$\left< P_T^{jet}\right> \approx \left< P_T^{\gamma ,Z}\right> $. It gives the 
attractive possibility to search for coherent effects in QCD-medium: the dependence of 
energy losses of the distance traversed can be studied experimentally in different bins 
of impact parameter distribution of nucleus-nucleus collision, or by varying collided 
ions and selecting the most central collisions. The intriguing prediction associated
with the coherence pattern of the medium induced radiation is that radiative energy 
losses per unit distance $dE/dx$ depend on the total distance traversed 
$L$~\cite{baier,baier2}. The value $dE/dx$ is approaching to being proportional to $L$ 
for static medium~\cite{baier}, and it has weaker $L$-dependence for the case of 
expanding medium~\cite{baier2}. The main goal of the present paper is to analyze the 
possibility of observing the $L$-dependence of jet energy losses $dE/dx$ for realistic 
nuclear geometry. In particular, we are studying the impact parameter dependence of 
collisional and radiative jet energy losses in dense QCD-matter, created in 
ultrarelativistic heavy ion collisions.    

\section {The geometrical model for jet production in nuclear collisions} 

Let us to consider the simple geometrical model of jet production and jet passing 
through a dense matter in high energy symmetric nucleus-nucleus collision. The figure 1 
shows the essence of the problem in the plane of impact parameter {\bf b} of two 
colliding nuclei $A$-$A$. The impact parameter $b$ here is the transverse distance 
between nucleus centers $O_1$ and $O_2$, $OO_2 = -O_1O = b/2$. Let $B(r\cos{\psi}, 
r\sin{\psi})$ be denoted as a jet (dijet) production vertex, with $r$ being the 
distance from the nuclear collision axis to the $B$. Then the distance between 
nucleus centers ($O_1, O_2$) and vertex $B$ can be found as  
\begin{equation} 
r_{1,2} = \sqrt{r^2+\frac{b^2}{4}\pm rb\cos \psi} . 
\end{equation} 

\begin{figure}[hbtp]
\centerline{\makebox{\epsfig{figure=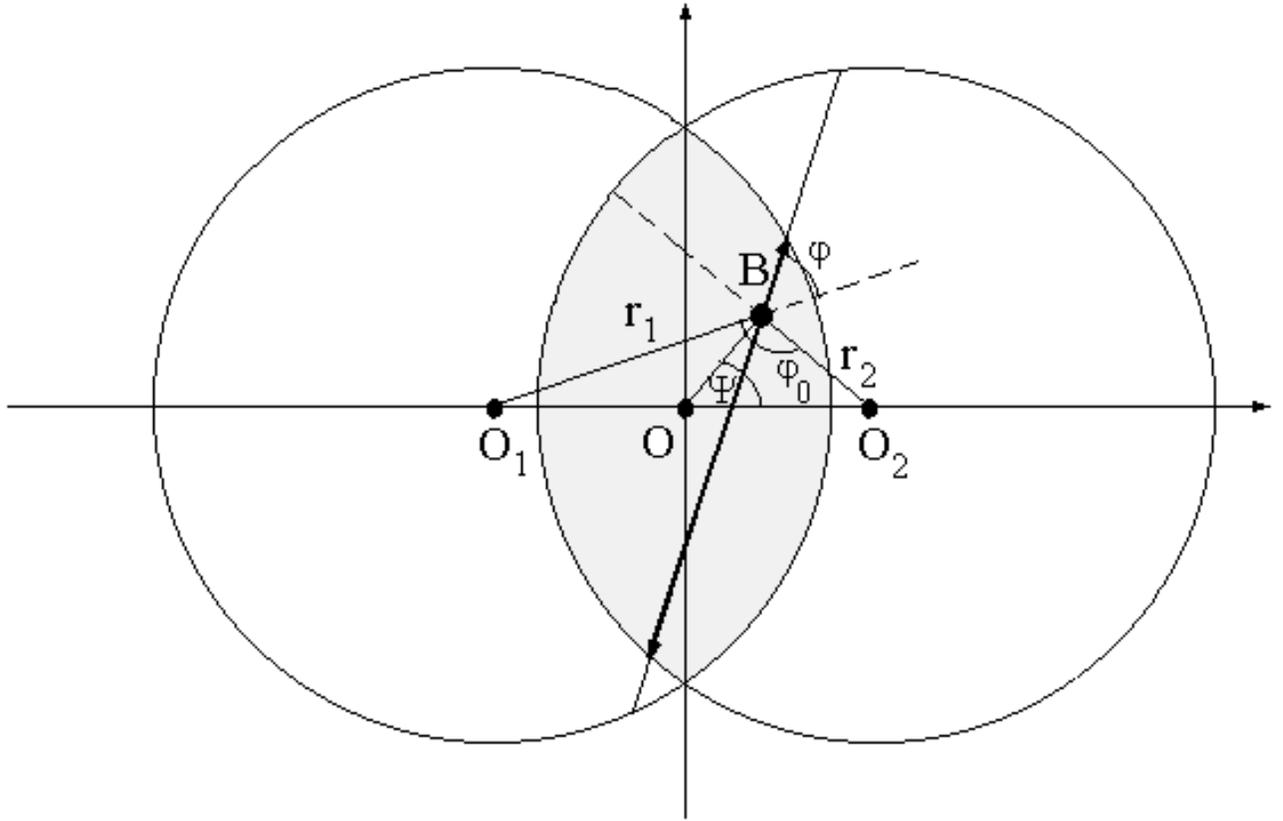,height=16cm}}} 
\caption{\label{fig1}  Jet production in high energy symmetric nucleus-nucleus 
collision in the plane of impact parameter {\bf b}. $O_1$ and $O_2$ are nucleus 
centers, $OO_2 = -O_1O = b/2$. $B(r\cos{\psi}, r\sin{\psi})$ is the jet (dijet) 
production vertex; $r$ is  the distance from the nuclear collision axis to $B$; 
$r_1,r_2$ are distances between nucleus centers ($O_1, O_2$) and $B$; $\varphi$ is the 
jet azimuthal angle; $\varphi_0$ is the azimuthal angle between vectors ${\bf r_1}$ 
and ${\bf r_2}$. }
\end{figure}

The distribution over jet production vertex $B (r, \psi)$ at given impact 
parameter $b$ is written as 

\begin{equation} 
\label{vertex}
P_{AA}({\bf r}, b) = \frac{T_A(r_1)\cdot T_A(r_2)}{T_{AA}(b)}, 
\end{equation} 
where 
\begin{equation} 
T_{AA}(b) = \int d^2 {\bf s} T_A ({\bf s}) T_A ({\bf b} - {\bf s}) = 
\int\limits_0^{2\pi} d\psi \int\limits_0^{r_{max}} rdrT_A(r_1)T_A(r_2) 
\end{equation} 
is the nuclear overlap function, $T_A({\bf r}) = A \int\limits_{-\infty}^{+\infty} 
\rho_A({\bf r},z)dz$ is the nuclear thickness function with nucleon density
distribution $\rho_A({\bf r},z)$. The maximum possible value of $r$ in nuclear 
overlapping zone can be estimated from the equation  
\begin{equation} 
max \{ r_1 (r=r_{max}), r_2 (r=r_{max})\} = R_A  
\end{equation} 
($R_A$ is the radius of the nucleus $A$). 
This gives 
\begin{equation} 
r_{max} = min\{ \sqrt{R_A^2 - \frac{b^2}{4} \sin^2 \psi} + \frac{b}{2} \cos 
\psi ,  \sqrt{R_A^2 - \frac{b^2}{4} \sin^2 \psi} - \frac{b}{2} \cos \psi \} . 
\end{equation} 
In particular, for the uniform nucleon density distribution, 
$\rho^{un}_A({\bf R}) = \rho_0 \cdot \Theta (R_A-|{\bf R}|)$, the nuclear overlap 
function is equal to $T^{un}_A(r) = 3A \sqrt{R_A^2-r^2} / (2 \pi R_A^3)$. Then the 
distribution $P^{un}_{AA}({\bf r}, b)$ is proportional to 
\begin{equation}  
P^{un}_{AA}({\bf r}, b) \propto \sqrt{R_A^2 - r_1^2(r,\psi,b)}\cdot 
\sqrt{R_A^2 - r_2^2(r,\psi,b)} . 
\end{equation} 
For central $AA$ collisions ($b=0$, $r_{max}=R_A$) we get simply $P^{un}_{AA}({\bf r},
b=0) \propto (R_A^2-r^2)$. 

It is straightforward to evaluate the time $\tau_L = L$ it takes for jet to traverse 
the dense zone:   
\begin{equation} 
\label{taul} 
\tau_L = min\{\sqrt{R_A^2 -r_1^2\sin^2\varphi} - r_1
\cos\varphi,~\sqrt{R_A^2 - r_2^2 \sin^2(\varphi-\varphi_0)} - r_2
\cos(\varphi-\varphi_0)\} ,  
\end{equation} 
where $\varphi$ is the azimuthal angle which determines the direction of a 
jet motion in the transverse plane, and $\varphi_0$ is the angle between vectors 
${\bf r_1}$ and ${\bf r_2}$. The expression for  
\begin{equation} 
\varphi_0 = \arccos{\frac{r^2-b^2/4}{r_1r_2}} 
\end{equation} 
can be obtained from the condition 
\begin{equation} 
r_1 r_2 \cos{\varphi_0} = {\bf r_1} \cdot {\bf r_2} = 
(-b/2-r\cos{\psi})\cdot(b/2-r\cos{\psi})+r^2 \sin^2{\psi} = r^2 -b^2/4 . 
\end{equation} 

Finally, we are going to estimate the dependence of initial energy density in nuclear 
overlapping zone on impact parameter of the collision. At collider energies the minijet 
system (the semi-hard gluons, quarks and antiquarks with $p_T\ga p_0 \sim 1 \div 2$ 
GeV/c) in the central rapidity region is typically formed in parton-parton scatterings 
at very early times, $\tau_0 \sim 1/p_T \la 1/p_0 \sim 0.1$ fm/c, and it will then 
serve as initial condition for the further evolution of the system~\cite{eskola94}. 
Strictly speaking, the soft particle production mechanisms (like the decay of the 
colour field) can also contribute to initial conditions in nuclear interactions. 
However, the relative strength of soft part decreases strongly with increasing c.m.s. 
energy of the ion beams. In particular, at LHC energies $\sqrt{s}=7$ TeV$\times (2Z/A)$ 
per nucleon pair the hard and semi-hard processes contribute over $80 \%$ to the 
transverse energy in heavy ion collisions~\cite{eskola94}. Moreover, soft processes 
with small momentum transfer $Q^2 \sim \Lambda^2_{QCD} \simeq (200$ MeV$)^2$ $\ll 
p_0^2$ can be partially or fully suppressed, owing to screening of the colour 
interaction in the dense parton matter produced from the system of minijets in the 
early stages of the reaction~\cite{eskola96}. Therefore, at LHC energies, we will 
consider only dominant semi-hard contribution to the formation of initial state. 

The initial energy density inside the comoving volume of longitudinal size $\Delta z = 
\tau_0\cdot 2\Delta y$ can be estimated using the Bjorken 
formula~\cite{bjorken,eskola94} as 
\begin{equation} 
\varepsilon (\tau = \tau_0) =  \frac{\left< E_T^A(|y|<\Delta y)\right> }{S(b)\cdot 
\Delta z} = \frac{\left< E_T^A(|y|<\Delta y)\right> \cdot p_0}{S(b)\cdot 2\Delta y} , 
\end{equation} 
where 
\begin{equation} 
S_{AA}(b) = \int\limits_0^{2\pi} d\psi \int\limits_0^{r_{max
}}rdr = \left( \pi -2 
\arcsin{\frac{b}{2R_A}} \right) R_A^2 - b\sqrt{R_A^2-\frac{b^2}{4}} 
\end{equation} 
is the effective transverse area of nuclear overlapping zone at impact parameter $b$. 
The total initial transverse energy deposition in mid-rapidity region can be 
calculated~\cite{eskola94} as 
\begin{equation} 
\left< E_T^A (b, \sqrt{s},p_0,|y|<\Delta y) \right> =
T_{AA}(b) \cdot \sigma_{NN}^{jet} (\sqrt{s},p_0) \cdot \left< p_T \right> ,  
\end{equation} 
where the first $p_T$-moment of inclusive differential minijet cross section 
$\sigma_{NN}^{jet} \cdot \left< p_T \right>$ is determined by the dynamics of 
nucleon-nucleon interactions at the corresponding c.m.s. energy. Then the dependence of 
initial energy density $\varepsilon_0$ in nuclear overlapping zone on impact parameter 
$b$ has the form: 
\begin{equation} 
\label{eps_1} 
\varepsilon _0 (b) \propto T_{AA}(b)/S_{AA}(b) ,  
\end{equation} 
or 
\begin{equation} 
\label{eps_0}
\varepsilon_0(b)=\varepsilon_0(b=0)\frac{T_{AA}(b)}{T_{AA}(b=0)}\frac{S_{AA}(b=0)}{S_{AA}(b)} 
.
\end{equation} 
For central $AA$ collisions we have $S_{AA}(b=0) = \pi R_A^2$ and $T_{AA}(b=0) 
= 9A^2 / (8 \pi R_A^2)$. 

It is worth noting that although this simple geometrical model for jet production in 
nucleus-nucleus collisions is formally can be applicable up to impact parameter 
$b = 2 R_A$, the major informative domain of our interest is central and semi-central 
collisions with $b \la R_A$ only. We have the following reasons in favour of this. 

1) The contribution of such events to total jet rate is dominant, although these events 
represent only a few percents of total inelastic $AA$ cross section~\cite{vogt99}. For 
example, the $Pb-Pb$ collisions with impact parameter $b < 0.9 R_{Pb} = 6$ fm 
contribute $\approx 50 \%$ to the total dijet rate at LHC energy, their relative 
fraction of total cross section being only $\approx 10 \%$ 
in this case~\cite{note99-016}. 

2) In the most central heavy ion collisions the maximum initial energy density is 
expected to be achieved in a fairly large (compared with typical hadronic scales) 
volume, when the effect of super-dense and hot matter formation, like quark-gluon 
plasma, can be really observable. The result for impact parameter dependence of initial 
energy density $\varepsilon_0$ (\ref{eps_0}) in nuclear overlapping zone for uniform 
nucleon density is shown in figure 2: it is very weakly dependent of $b$ ($\delta 
\varepsilon_0 \la 10 \%$) up to $b \sim R_A$, and decreases rapidly at $b \ga R_A$. On 
the other hand, the averaged over all possible jet production vertices proper time 
$\left< \tau_L \right> $ (\ref{taul}) of jet escaping from the dense zone is found to 
go down almost linearly with increasing impact parameter $b$ (see the second curve in 
fig.2). Therefore the variation of impact parameter $b$ of nucleus-nucleus collision 
(which can be measured, for example, using the total transverse energy deposition
detected in different parts of calorimeters~\cite{note99-015}) up to $b \sim R_A$ gives 
the possibility to study jet quenching as a function of distance traversed without 
significant changing initial energy density $\varepsilon_0$. 

Meanwhile, the weakness of $b$-dependence of $\varepsilon_0$ gives us the advantage
as compared with using of beams of different ions at a fixed bin of impact parameter
distribution, when the scaling $\left< \tau_L \right> (b=0) \propto R_A \propto 
A^{1/3}$ exists. Eq.(\ref{eps_1}) gives $\varepsilon_0 (b=0) \propto A^2/R_A^4$, i.e. 
$\varepsilon_0 (b=0) \propto A^{2/3}$. Figure 3 illustrates changing of average
time $\left< \tau_L \right>$ of jet travel and initial energy density $\varepsilon_0$ 
in dense zone with variation of impact parameter $b$ (at fixed $A=207$, Pb) and atomic
weight $A$ (at fixed $b=0$). For example, decreasing $\left< \tau_L \right> (Pb,b=0) 
\simeq 6$ fm by the factor $\simeq 1.7$ can be obtained by:\\ a) increasing $b$ up to
$b=0.9R_{Pb}\simeq 6$ fm (at the expense of only $\sim 10 \%$ of $\varepsilon_0$ 
reduction); \\ b) decreasing $A$ down to $A=40$ (Ca) (at the expense of $\varepsilon_0$ 
reduction by the factor $\sim 3$). 

3) It is well known, that the uniform nucleon density distribution in the nucleus, 
$\rho^{un}_A({\bf R}) = \rho_0 \cdot \Theta (R_A-|{\bf R}|)$, can serve as a good 
approximation for central and semi-central collisions (see figure 4, which shows the 
nuclear overlap function profile for the uniform~\footnote{Moreover, in this case the
explicit form of $T_{AA} (b)$ can be obtained: 
$$T^{un}_{AA}(b) = T^{un}_{AA}(b=0) \left[ 1 - \tilde{b} \left[ 1+\left( 
1-\frac{\tilde{b}}{4}\right) \ln{\frac{1}{\tilde{b}}}+2\left( 1-\frac{\tilde{b}}{4} 
\right) \left( \ln{(1+\sqrt{1-\tilde{b}})}-\frac{\sqrt{1-\tilde{b}}}{1+
\sqrt{1-\tilde{b}}} \right) - \frac{\tilde{b}(1-\tilde{b})}{2(1+\sqrt{1-\tilde{b}})^2} 
\right] \right] , $$ $\tilde{b} = b^2 / (4 R_A^2)$, the weak $b$-dependence of 
$\varepsilon_0$ and approximately linear drop of $\left< \tau_L \right> (b)$ being 
derived for $b \la R_A$ analytically.} and the standard Woods-Saxon nucleon 
densities). The edge effects near the surface of the nucleus, impact parameter 
dependence of nuclear parton structure functions ("nuclear shadowing")~\cite{b-shad}, 
early transverse expansion of the system and other potentially important phenomena for 
peripheral ($b\sim 2R_A$) collisions are beyond our consideration here. 

\begin{figure}[hbtp]
\centerline{\makebox{\epsfig{figure=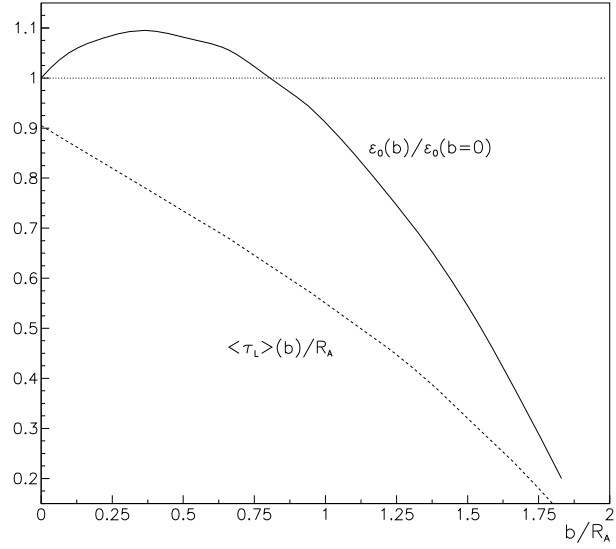,height=9cm}}} 
\caption{\label{fig2}  The impact parameter dependence of initial energy density 
$\varepsilon_0 (b) /  \varepsilon_0 (b=0)$ in nuclear overlapping zone (solid curve), 
and the average proper time $\left< \tau_L \right> / R_A$ of jet escaping from the 
dense matter (dashed curve) for uniform nucleon density distribution.}
\end{figure}

\begin{figure}[hbtp]
\centerline{\makebox{\epsfig{figure=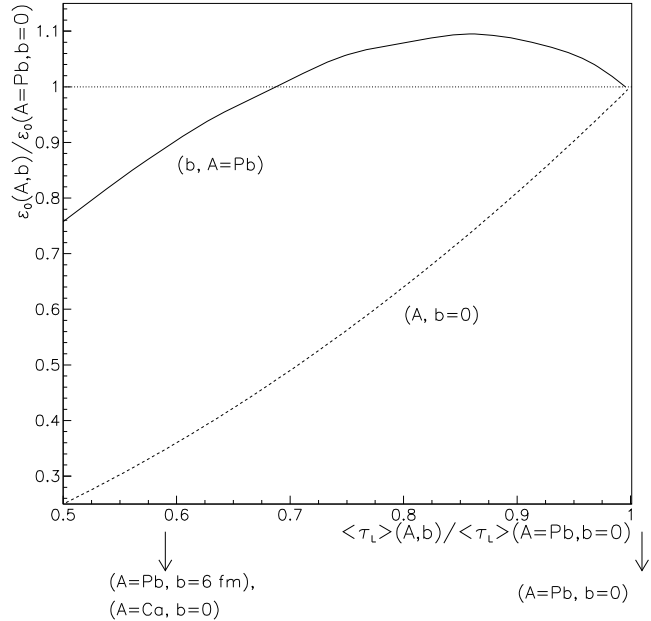,height=9cm}}} 
\caption{\label{fig3} The initial energy density $\varepsilon_0 (A,b) /  \varepsilon_0 
(A=Pb,b=0)$ in nuclear overlapping zone versus average proper time $\left< \tau_L 
\right> (A,b) / \left< \tau_L \right> (A=Pb,b=0)$ of jet escaping from the dense matter 
for varying atomic weight $A$ at fixed $b=0$ (dashed curve), and impact parameter $b$ 
at fixed $A=Pb=207$ (solid curve) for uniform nucleon density distribution.} 
\end{figure}

\begin{figure}[hbtp]
\centerline{\makebox{\epsfig{figure=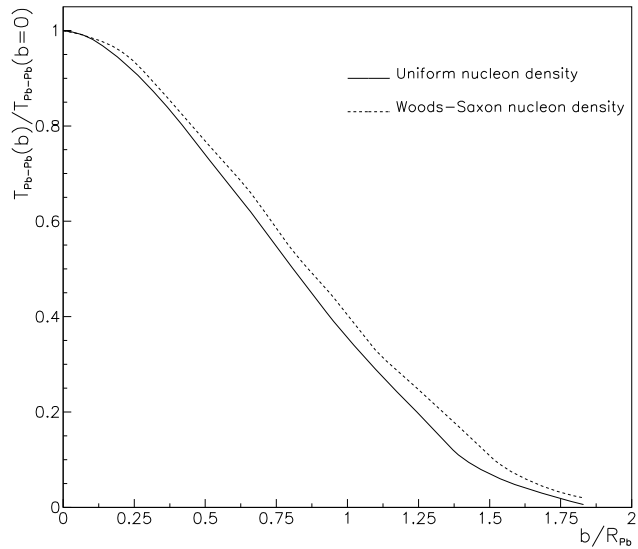,height=9cm}}} 
\caption{\label{fig4} The nuclear overlap profile function $T_{AA}(b)/T_{AA} (b=0)$ for 
uniform (solid) and Woods-Saxon (dashed) nucleon densities in Pb-Pb collisions.}  
\end{figure}

\section{Impact parameter dependence of jet energy losses} 

The intensity of final state rescattering and collisional and radiative energy losses
of hard jet partons in dense QCD-matter, created in nuclear overlapping zone, are 
sensitive to their initial parameters (energy density, formation time) and space-time
evolution~\cite{lokhtin1}. In order to analyze the impact parameter dependence of jet 
energy losses and jet quenching, we treat the medium as a boost-invariant 
longitudinally expanding quark-gluon fluid, and partons as being produced on a 
hyper-surface of equal proper times $\tau = \sqrt{t^2 -  z^2}$~\cite{bjorken}.   
We are expecting this is a adequate approximation for central and semi-central
collisions for our semi-qualitative discourse.  
 
The approach relies on an accumulative energy losses, when both initial and final state 
gluon radiation is associated with each scattering in expanding medium together 
including the interference effect by the modified radiation spectrum as a function of 
decreasing temperature $dE/dx(T)$. Note that recently the radiative energy losses of a 
fast parton propagating through expanding (according to Bjorken's model) QCD plasma 
have been explicitly evaluated in~\cite{baier2} as $dE / dx | _{expanding} = c \cdot 
dE / dx | _{T_L}$ with numerical factor $c \sim 2$ $(6)$ for a parton created inside 
(outside) the medium, $T_L$ being the temperature at which the dense matter was 
left~\cite{baier2}. 

The total energy losses in transverse direction experienced by a hard parton due to 
multiple scattering in matter are the result of averaging over the jet production 
vertex $P_{AA}({\bf r},b)$ (\ref{vertex}), the transfer momentum squared $t$ in a 
single rescattering and space-time evolution of the medium:  
\begin{equation}
\label{en_los} 
\left< \Delta E_T (b)\right> = \int\limits_0^{2\pi} d \psi 
\int\limits_0^{r_{max}}r\cdot dr  \frac{T_A(r_1)\cdot T_A(r_2)}{T_{AA}(b)}
\int\limits_0^{2\pi}\frac{d\varphi}{2\pi}
\int\limits_{\displaystyle\tau_0}^{\displaystyle 
\tau_L}d\tau \left( \frac{dE}{dx}^{rad}(\tau) + \sum_{b}\sigma_{ab}(\tau)\cdot
\rho_b(\tau)\cdot \nu(\tau) \right) .    
\end{equation} 
Here $\tau_0$ and $\tau_L$ (\ref{taul}) are the proper time of the plasma formation and 
the time of jet escaping from the dense zone respectively; $\rho_b \propto T^3$ is the 
density of plasma constituents of type $b$ at temperature $T$; $\sigma_{ab}$ is the 
integral cross section of scattering of the jet parton $a$ off the comoving constituent 
$b$ (with the same longitudinal rapidity $y$); $\nu$ and $dE / dx ^{rad}$ are 
the thermal-averaged collisional energy loss of a jet parton due to single elastic 
scattering and radiative energy losses per unit distance respectively.  

If the mean free path of a hard parton is larger than the screening radius in the QCD 
medium, $\lambda \gg \mu_D^{-1}$, the successive scatterings can be treated as 
independent~\cite{gyul94}. The transverse distance between successive scatterings, 
$\Delta r_i = (\tau_{i+1} - \tau_i) \cdot v_T = (\tau_{i+1} - \tau_i) \cdot p_T/E$, 
is determined in linear kinetic theory according to the probability density:  
\begin{equation} 
\frac{dP}{d(\Delta r_i)} = \lambda^{-1}(\tau_{i+1})\cdot 
\exp{(-\int\limits_0^{\Delta r_i}\lambda^{-1} (\tau_i + s)ds)},
\end{equation}
where the mean inverse free path is given by
$\lambda_a^{-1}(\tau) = \sum_{b}\sigma_{ab}(\tau) \rho_b(\tau)$. 

The dominant contribution to the differential cross section $d\sigma / dt$ for 
scattering of a parton with energy $E$ off the "thermal" partons with energy (or 
effective mass) $m_0 \sim 3T \ll E$ at temperature $T$ can be written 
as~\cite{gyul94,thoma98}  
\begin{equation} 
\frac{d\sigma_{ab}}{dt} \cong C_{ab} \frac{2\pi\alpha_s^2(t)}{t^2}, 
\end{equation} 
where $C_{ab} = 9/4, 1, 4/9$ for $gg$, $gq$ and $qq$ scatterings respectively,  
\begin{equation} 
\alpha_s = \frac{12\pi}{(33-2N_f)\ln{(t/\Lambda_{QCD}^2)}} \>   
\end{equation} 
is the QCD running coupling constant for $N_f$ active quark flavours, and 
$\Lambda_{QCD}$ is the QCD scale parameter which is of the order of the critical 
temperature,  $\Lambda_{QCD}\simeq T_c$. The integrated parton scattering cross 
section, 
\begin{equation} 
\sigma_{ab} = \int\limits_{\displaystyle
\mu^2_D(\tau)}^{\displaystyle m_0(\tau)E / 2 }dt\frac{d\sigma_{ab}}{dt}\>, 
\end{equation} 
is regularized by the Debye screening mass squared $\mu_D^2$. 

The collisional energy losses due to elastic scattering with high-momentum 
transfer have been originally estimated by Bjorken in~\cite{bjork82}, and recalculated
later in~\cite{mrow91} taking also into account the loss with low-momentum transfer 
dominated by the interactions with plasma collective modes. Since latter process 
contributes to the total collisional energy losses without the large factor 
$\sim \ln{(E / \mu_D)}$ in comparison with high-momentum scattering and it can be 
effectively "absorbed" by the redefinition of minimal $t \sim \mu_D^2$ 
under the numerical estimates, we shall concentrate on collisional energy losses with 
high-momentum transfer only~\footnote{Anyway, high- and low-momentum parts of
collisional energy losses have the same dependence on distance traversed.}. The thermal 
average of such loss can be written as
\begin{equation}  
\label{nu_col}
\nu = \lrang{\frac{t}{2m_0}} = \frac12\lrang{\frac1{m_0}}\cdot\lrang{t} 
\simeq \frac{1}{4T \sigma_{ab}} 
\int\limits_{\displaystyle\mu^2_D}^
{\displaystyle 3T E / 2}dt\frac{d\sigma_{ab}}{dt}t \> .
\end{equation} 
The value $\nu$ is independent of total distance traversed and determined by 
temperature, roughly $\nu \propto T$. Then total collisional energy losses integrated 
over whole jet path are estimated as $\left< \Delta E_{col} \right> \propto T_0^2 
\propto \sqrt{\varepsilon_0}$, as it has been pointed out in~\cite{bjork82}. 
The $\tau_L$-dependence of $\Delta E_{col}$ can be weaker than linear for expanding 
medium ($\Delta E_{col} \propto \tau_L$ for static matter). 

The energy spectrum of coherent medium-induced gluon radiation and the corresponding
dominated part of radiative energy losses, $dE/dx$, were analyzed 
in~\cite{baier,baier2} by means of the Schr\"odinger-like equation whose "potential" is 
determined by the single-scattering cross section of the hard parton in the medium. For 
the quark produced in the medium it gives~\cite{baier2,zakharov}~
\footnote{The gluons with formation times $\tau_f$ exceeding the time $\tau_L=L$ that 
are formed outside the medium (the factorization {\em medium-independent}\/ component) 
carry away a fraction of the initial parton energy proportional to $\alpha_s(E)$. This 
part of gluon radiation produces the standard jet energy profile which is identical to 
that of a jet produced in a hard process in the vacuum. Hereafter we shall concentrate 
on the {\em medium-dependent}\/ effects and will not include the "vacuum" part of the 
jet profile.}   
\begin{eqnarray} 
\label{radiat} 
\frac{dE}{dx}^{rad} = \frac{2 \alpha_s C_R}{\pi \tau_L}
\int\limits_{\omega_{\min}}^E  
d \omega \left[ 1 - y + \frac{y^2}{2} \right] 
\>\ln{\left| \cos{(\omega_1\tau_1)} \right|} 
\>, \\  
\omega_1 = \sqrt{i \left( 1 - y + \frac{C_R}{3}y^2 \right)   
\bar{\kappa}\ln{\frac{16}{\bar{\kappa}}}}
\quad \mbox{with}\quad 
\bar{\kappa} = \frac{\mu_D^2\lambda_g  }{\omega(1-y)}.
\end{eqnarray} 
Here $\tau_1 = \tau_L / (2 \lambda_g)$, and $y = \omega / E$ is the fraction of the 
hard parton energy carried by the radiated gluon, and $C_R = 4/3$ is the quark colour 
factor. A similar expression for the gluon jet can be obtained by substituting 
$C_R=3$ and a proper change of the factor in the square bracket in (\ref{radiat}), 
see~\cite{baier2}. The integral (\ref{radiat}) is carried out over all energies from 
$\omega_{\min}=E_{LPM}=\mu_D^2\lambda_g$ ($\lambda_g$ is the gluon mean free path), 
the minimal radiated gluon energy in the coherent LPM regime, up to initial jet energy 
$E$. The complex form of the expression (\ref{radiat}) does not allow us in general 
case to extract the explicit form of $\tau_L$- and $T$- dependences of 
${dE}/{dx}^{rad}$. In the limit of "strong" LPM effect, $\omega \gg \mu_D^2\lambda_g$, 
we have~\cite{baier,baier2,baier3} ${dE}/{dx}^{rad} \propto T^3$ and ${dE}/{dx}^{rad} 
\propto \tau_L$ with logarithmic accuracy. Then total radiative energy losses 
$\left< \Delta E_{rad} \right> = \int d\tau \cdot {dE}/{dx}^{rad}$ are estimated 
as $\left< \Delta E_{rad} \right> \propto T_0^3 \propto \varepsilon_0^{3/4}$ and 
$\Delta E_{rad} \propto \tau_L^{\beta}$, where $\beta \la 2$ for expanding medium 
($\beta \sim 2$ in the case of static matter). 
 
In order to simplify numerical calculations (and not to introduce new parameters) we 
omit the transverse expansion and viscosity of the fluid using the well-known scaling 
Bjorken's solution~\cite{bjorken} for temperature and density of QGP at $T > T_c \simeq 
200$ MeV: 
\begin{equation} 
\varepsilon(\tau) \tau^{4/3} = \varepsilon_0 \tau_0^{4/3},~~ 
T(\tau) \tau^{1/3} = T_0 \tau_0^{1/3},~~ \rho(\tau) \tau = \rho_0 \tau_0 . 
\end{equation}
Let us remark that the influence of the transverse flow, as well as of the mixed phase 
at $T = T_c$, on the intensity of jet rescattering (which is a strongly increasing 
function of $T$) seems to be inessential for high initial temperatures 
$T_0 \gg T_c$~\cite{lokhtin1}. On the contrary, the presence of viscosity slows down 
the cooling rate, which leads to a jet parton spending more time in the hottest regions 
of the medium. As a result the rescattering intensity goes up, i.e., in fact an 
effective temperature of the medium gets lifted as compared with the perfect QGP 
case~\cite{lokhtin1}. We also do not take into account here the probability of jet 
rescattering in nuclear matter, because the intensity of this process and corresponding 
contribution to total energy losses are not significant due to much smaller energy 
density in a "cold" nuclei. For certainty we used the initial conditions for the 
gluon-dominated plasma formation ($N_f \approx 0$, $\rho_q \approx 1.95T^3$) expected 
for central $Pb-Pb$ collisions at LHC~\cite{eskola94}: $\tau_0 \simeq 0.1$ fm/c, $T_0 
\simeq 1$ GeV~\footnote{These estimates are of course rather 
approximate and model-depending: the discount of higher order $\alpha_s$ terms, 
uncertainties of structure functions in the low-$x$ region, and nuclear shadowing can 
result in variations of the initial energy density~\cite{eskola94}.}. 

Figure 5 represents the calculated $\tau_L$-dependence of coherent medium-induced  
radiative and collisional energy losses of a quark-initiated jet with initial energy 
$E_T^q = 100$ GeV. We see that the $\tau_L$-dependence of radiative and collisional 
losses is very different: $\Delta E_{rad} (\tau_L)$ grows somewhat stronger than 
linearly, meanwhile $\Delta E_{col} (\tau_L)$ looks rather logarithmic. This results in 
the corresponding difference in the impact parameter dependence of radiative and 
collisional losses, the normalized profile of which are presented in figure 6. To make 
the plot more visual the energy losses $\left< \Delta E_T (b)\right>$ are normalized to 
the corresponding average values at zero impact parameter, $\left< \Delta E^q_{T~rad} 
(b=0)\right> \sim 45$ GeV and $\left< \Delta E^q_{T~col} (b=0)\right> \sim 5$ GeV for 
the parameters used. 

For example, decreasing of impact parameter from 
$b=0$ to $b=R_{Pb}$ gives $\sim 30 \%$ collisional and $\sim 50 \%$ radiative losses 
reduction. We have also found that the the form of $b$-dependence of collisional losses 
is almost independent of scenarios of space-time evolution of QGP (perfect or viscous 
fluid), $b$-dependence of radiative losses being somewhat more sensitive to these 
effects. 

\begin{figure}[hbtp]
\centerline{\makebox{\epsfig{figure=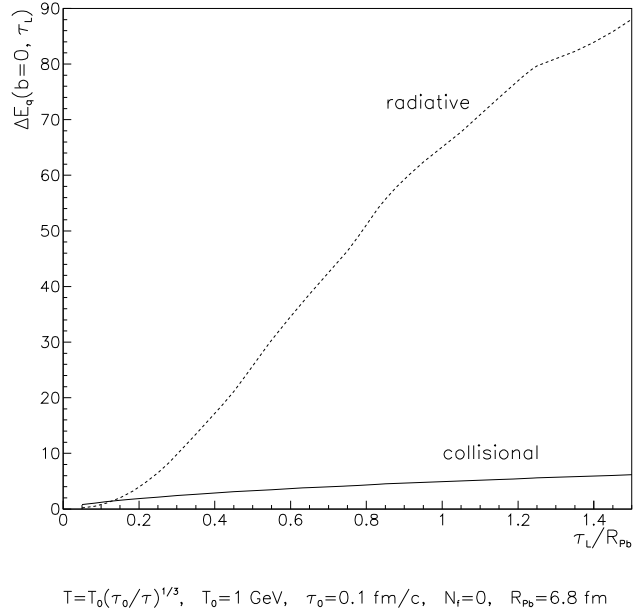,height=9cm}}} 
\caption{\label{fig5} The medium-induced radiative (dashed) and collisional (solid) 
energy losses of a quark-initiated jet with initial energy $E_T^q = 100$ GeV versus the 
average proper time $\left< \tau_L \right> / R_{Pb}$ of jet escaping from the dense 
matter.} 
\end{figure}

\begin{figure}[hbtp]
\centerline{\makebox{\epsfig{figure=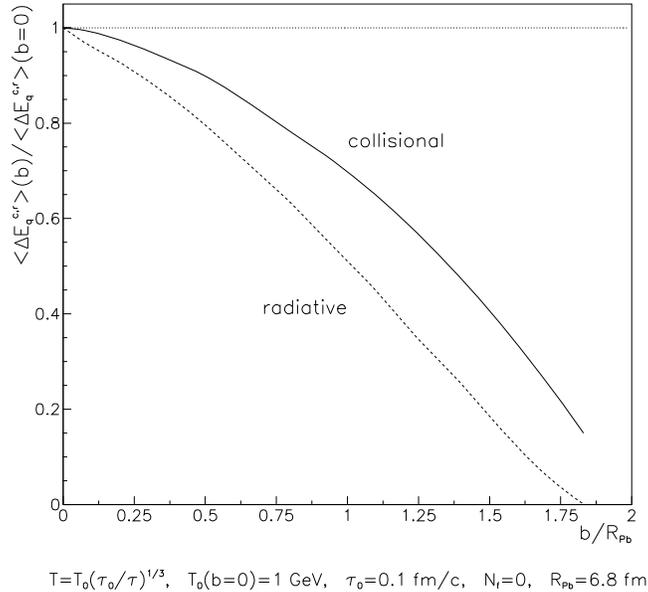,height=9cm}}} 
\caption{\label{fig6} The impact parameter dependence of medium-induced radiative 
(dashed) and collisional (solid) energy losses of a quark-initiated jet with initial 
energy $E_T^q = 100$ GeV normalized to the corresponding average values at zero impact 
parameter.} 
\end{figure} 

Note that the choice of the scale for a minimal jet energy $E_T^q \sim 100$ GeV 
corresponds to the threshold for "true" QCD-jet recognition against the "thermal" 
background jets (statistical fluctuations of the transverse energy flux) with 
reconstruction efficiency closed to $1$ in heavy ion collisions at 
LHC~\cite{note99-016,cms94,lokhtin3}. We hope that the separation of collisional 
and radiative contribution to the total energy losses, when doing the experimental data 
analysis for jets with finite cone size, could be performed basing on essential 
difference of their angular distribution~\cite{lokhtin2,baier3}: the radiative losses 
are expected to dominate at small jet cone size $\theta_0$, while the relative 
contribution to collisional losses grows with increasing $\theta_0$. 

\section{Impact parameter dependence of dijet production rate}

In previous section we have analyzed the impact parameter dependence of jet energy
losses, which can be directly observed in $\gamma + jet$ and $Z(\rightarrow \mu^+ 
\mu^-) + jet$ production processes. Another observable effect is a suppression of 
high-$p_T$ jet pair yield (dijet quenching) due to final state rescattering and energy 
losses. In connection with this, we would like to estimate the impact parameter 
dependence of $jet + jet$ production rates in heavy ion collisions. The observed number 
of $\{ij\}$ type dijets with transverse momenta $p_{T1}, p_{T2}$ produced in initial 
hard scattering processes in minimum bias $AA$ collisions is written as: 
\begin{eqnarray} 
\frac{dN_{ij}^{dijet}}{dp_{T1}dp_{T2}} & = & \int\limits_0^\infty d^2b \frac{d^2 
\sigma^0_{jet}}{d^2b} \cdot \frac{dN_{ij}^{dijet}}{dp_{T1}dp_{T2}} (b) \Bigg/ 
\int\limits_0^\infty d^2b \frac{d^2 \sigma^0_{jet}}{d^2b} , \\ 
\frac{dN_{ij}^{dijet}}{dp_{T1}dp_{T2}} (b) & = & \int\limits_0^{2\pi} d \psi 
\int\limits_0^{r_{max}}r dr T_A(r_1) T_A(r_2) 
\int\limits_0^{2\pi}\frac{d\varphi}{2\pi}
\int dp_T^2\frac{d\sigma_{ij}} {dp_T^2}~\delta(p_{T1} - p_T + \nonumber \\  
 & & \Delta E_T^i(r, \psi, \varphi, b))~ 
\delta(p_{T2} - p_T + \Delta E_T^j(r, \psi, \pi - \varphi, b)),
\end{eqnarray} 
where parton differential cross section $d\sigma_{ij} / dp_T^2$ is calculated in the 
perturbative QCD: 
\begin{equation}
\label{hardsec}
\frac{d\sigma_{ij}}{dp_T^2} = K \int dx_1 \int dx_2 \int 
d\widehat{t} f_i(x_1, p^2_T) f_j(x_2, p^2_T) 
\frac{d\widehat{\sigma}_{ij}}{d\widehat{t}} \delta 
(p^2_T - \frac{\widehat{t} \widehat{u}}{\widehat{s}}) ,   
\end{equation}
$d\widehat{\sigma_{ij}} / d\widehat{t}$ expresses the differential cross-section for a
parton-parton scattering as a function of the kinematical Mandelstam variables 
$\widehat{s}$, $\widehat{t}$ and $\widehat{u}$, $f_{i,j}$ are the structure functions, 
$x$ is the nucleon-momentum fraction carried by a parton, the correction factor $K$ 
takes into account higher order contributions. We have tested with the program of 
S.D.Ellis et al.~\cite{ellis} that next-to-leading order (NLO) corrections are  
insignificant ($K \sim 1$) for jets with $p_T \ge 50\div 100$ GeV/c and reasonable cone 
radius in the ($y,\phi$)-plane $R=0.3\div 0.5$ (see also ~\cite{eskola95}). Note also 
that the region of sufficiently hard jets, $x_{1,2} \sim \sqrt{\widehat{s}/s} \ga 0.2$, 
almost does not affected by the initial state nuclear interactions like gluon depletion 
("nuclear shadowing" of nucleon structure functions)~\cite{eks}. Anyway, the integrated 
above the threshold value $p_T^{cut}$ dijet rate, 
\begin{equation} 
\label{dijet_rate} 
R^{dijet}_{AA}(p_{T1}, p_{T2} > p_T^{cut}) = \int\limits_{p_T^{cut}}
dp_{T1}\int\limits_{p_T^{cut}}dp_{T2}\sum_{i,j}(\frac{\displaystyle dN_{ij}^{dijet}}
{\displaystyle dp_{T1}dp_{T2}})_{AA} , 
\end{equation} 
in $AA$ relative to $pp$ collisions can be studied by introducing a reference process, 
unaffected by energy losses and with a production cross section proportional to the 
number of nucleon-nucleon collisions, such as Drell-Yan dimuons or (suitable for
LHC~\cite{kvat95}) Z$(\rightarrow \mu^+\mu^-)$ production,  
\begin{equation}
\label{dijet_nrate}  
R^{dijet}_{AA} / R^{dijet}_{pp} = \left( \sigma_{AA}^{dijet} / 
\sigma_{pp}^{dijet} \right) /  \left(\sigma_{AA}^{DY~(Z)} / \sigma_{pp}^{DY~(Z)} 
\right).   
\end{equation} 
The cross section $d^2 \sigma^0_{jet} / d^2b$ for initially produced jets in 
$AA$ collisions at given $b$ can be written as~\cite{note99-016,vogt99}: 
\begin{equation} 
\label{jet_prob}
\frac{d^2 \sigma^0_{jet}}{d^2b} ({\bf b}, \sqrt{s}) =  T_{AA} ({\bf b}) 
 \sigma _{NN}^{jet} (\sqrt{s}) \frac {d^2 \sigma^{AA}_{in}}{d^2b} ({\bf b}, \sqrt{s}) , 
\end{equation} 
where nucleon-nucleon collision cross section of the hard process $\sigma _{NN}^{jet}$ 
has been computed with PYTHIA model~\cite{pythia}. The differential 
inelastic $AA$ cross section is calculated as: 
\begin{equation} 
\frac {d^2 \sigma^{AA}_{in}}{d^2b} ({\bf b}, \sqrt{s}) = \left[ 1 - \left( 1- \frac{1}  
{A^2}T_{AA}({\bf b}) \sigma^{in}_{NN} (\sqrt{s}) \right) ^{A^2} \right]   
\end{equation} 
with inelastic non-diffractive nucleon-nucleon cross section 
$\sigma^{in}_{NN}$ ($\simeq 60$ mb for $\sqrt{s} = 5.5$ TeV).

\begin{figure}[hbtp]
\centerline{\makebox{\epsfig{figure=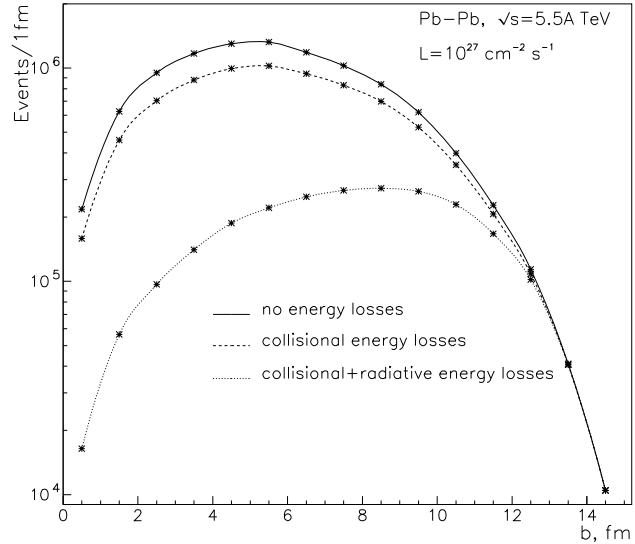,height=9cm}}} 
\caption{\label{fig7} The $jet+jet$ rates for $E_T^{jet} > 100$ GeV and $|y^{jet}| < 
2.5$ in different impact parameter bins for cases: without energy losses (solid curve), 
with collisional losses only (dashed curve), with collisional and radiative losses 
(dotted curve). The rates are normalized to the expected number of events produced in 
Pb-Pb collisions during two weeks of LHC running.} 
\end{figure}

Figure 7 shows dijet rates $\sigma^{in}_{AA} R^{dijet}_{AA} L$ ($E_T^{jet} > 
p_T^{cut} = 100$ GeV, rapidity window $|y^{jet}| < 2.5$) in different impact parameter 
bins for three cases: $(i)$ without energy losses, $(ii)$ with collisional losses
only, $(iii)$ with collisional and radiative losses. The rates are normalized to the 
expected number of events produced in Pb-Pb collisions during two weeks ($1.2 \times 
10^6$ s) of LHC run time, assuming luminosity $L = 10^{27}~cm^{-2}s^{-1}$~\cite{cms94}. 
The total initial dijet rate with $E_T^{jet} > 100$ GeV is estimated as $1.1 \times 
10^{7}$ events ($gg \rightarrow gg$ $\simeq 60 \%$, $qg \rightarrow qg$ $\simeq 30 \%$, 
$qq,gg \rightarrow qq$ $\simeq 10 \%$). Since the dijet quenching is much stronger in 
central collisions than in peripheral one's, the maximum and mean values of 
$dN^{dijet}/db$ distribution get shifted towards the larger $b$. The corresponding 
result for jets with non-zero cone size $\theta_0$ is expected to be somewhere between 
$(iii)$ ($\theta_0 \rightarrow 0$) and $(ii)$ cases. The observation of a dramatic 
change in the $b$-dependence of dijet rates in heavy ion collisions as compared to what 
is expected from independent nucleon-nucleon interactions pattern, would indicate the 
existence of medium-induced parton rescattering. 

As we have mentioned above, the measurement of the centrality of events can be 
performed from total transverse energy deposition $E^{tot}_T$ in calorimeters, which 
strongly decreases from central to peripheral collisions~\cite{note99-015}, roughly 
as $\overline{E_T^{tot}} (b) \propto T_{AA} (b)$. If jet energy losses 
$\left< \Delta E_T^{jet}\right> $ (\ref{en_los}) or dijet production 
rates $R^{dijet}$ (\ref{dijet_rate},\ref{dijet_nrate}) are measured in different bins 
of $E^{tot}_T$, then one can relate $b$- and $E^{tot}_T$- dependences of $F = (\Delta 
E_T^{jet}, R^{dijet})$ using $E^{tot}_T-b$ correlation functions $C_{AA}$: 
\begin{eqnarray} 
F (E^{tot}_T) = \int d^2b F(b) C_{AA} (E_T^{tot}, b),  
~~ C_{AA} (E_T^{tot}, b) = \frac{1}{\sqrt{2\pi}\sigma_{E_T}(b)} \exp {\left( - 
\frac{\left( E_T^{tot}-\overline{E_T^{tot}}(b) \right)^2}{2 \sigma_{E_T}^2(b)}\right) } 
, \\  F (b) = \int d E_T^{tot} F(E_T^{tot}) C_{AA} (b, E_T^{tot}),  
~~ C_{AA} (b, E_T^{tot}) = \frac{1}{\sqrt{2\pi}\sigma_{b}(E_T^{tot})} \exp {\left( - 
\frac{\left( b-\overline{b}(E_T^{tot}) \right)^2}{2 \sigma_{b}^2(E_T^{tot})}\right) } . 
\end{eqnarray} 
The estimated with the HIJING model~\cite{hijing} accuracy of impact parameter 
determination $\sigma_b (E^{tot}_T) \sim 1-2$ fm in $AA$ collisions at
LHC~\cite{shmatov} seems to be enough to observe the above effects.

\section{Conclusions}  

To summarize, we have considered the impact parameter dependence of medium-induced 
radiative and collisional jet energy losses in dense QCD-matter, created in 
ultrarelativistic heavy ion collisions. We have found that this $b$-dependence is    
very different for each mechanism due to coherent effects (the dependence of radiative 
energy losses per unit distance $dE/dx$ of total distance traversed). As a consequence, 
the radiative losses are more sensitive to the impact parameter of nucleus-nucleus 
collision, which determines the effective volume of nuclear overlapping dense zone,  
and the space-time evolution of the medium. 

A possible way to directly observe the energy losses at different impact parameter (or 
total detected $E_T$ deposition) bins, involves tagging the hard jet opposite a 
particle that does not interact strongly, like in $\gamma + jet$ and $Z(\rightarrow 
\mu^+ \mu^-) + jet$ production processes. Since initial energy density $\varepsilon_0$ 
in dense zone depends on $b$ very slightly ($\delta \varepsilon_0 \la 10 \%$) up to 
$b \sim R_A$, studying $b$-dependence appears to be advantageous than using of
different ions at fixed impact parameter $b \sim 0$ (when $\varepsilon_0 (b \sim 
0) \propto A^{2/3}$). We hope that the separation of collisional and radiative 
contribution to total energy losses when doing the experimental data analysis for jets 
with finite cone size could be performed basing on essential difference in their 
angular distributions. 

Another process of interest is high-$p_T$ jet pair production. The expected 
statistics for dijet rates in heavy ion collisions at LHC will be large 
enough to study the impact parameter dependence. Since suppression of dijet yield (jet 
quenching) due to medium-induced energy losses should be much stronger in 
central collisions than in the peripheral one's, the maximum and mean values of 
$dN^{dijet}/db$ distribution predicted to be shifted towards the larger $b$.

Finally, the study of the impact parameter dependences in the hard jet production 
processes ($jet + jet$, $\gamma + jet$ and $Z + jet$ channels) is important for 
extracting information about the properties of super-dense QCD-matter to be created in 
heavy ion collisions at LHC. 

$Acknowledgements$. Discussions with Yu.L.Dokshitzer, L.I.Sarycheva and R.Vogt are
gratefully acknowledged.

\end{document}